\documentclass[12pt]{article}
\usepackage{epsfig,graphicx}

\title{Relativistic path integral and relativistic Hamiltonians in QCD and QED}
\author{  Yu.A.Simonov,\\ Institute of Theoretical and Experimental
Physics\\ 117118, Moscow, B.Cheremushkinskaya 25, Russia}
\date{}

\newcommand{\be}{\begin{equation}}
\newcommand{\ee}{\end{equation}}

\def\la{\mathrel{\mathpalette\fun <}}
\def\ga{\mathrel{\mathpalette\fun >}}
\def\fun#1#2{\lower3.6pt\vbox{\baselineskip0pt\lineskip.9pt
\ialign{$\mathsurround=0pt#1\hfil ##\hfil$\crcr#2\crcr\sim\crcr}}}

\newcommand{{\SD}}{\rm SD}

\newcommand{\vex}{\mbox{\boldmath${\rm x}$}}
\newcommand{\vey}{\mbox{\boldmath${\rm y}$}}
\newcommand{\ver}{\mbox{\boldmath${\rm r}$}}
\newcommand{\vesig}{\mbox{\boldmath${\rm \sigma}$}}

\newcommand{\veP}{\mbox{\boldmath${\rm P}$}}
\newcommand{\vep}{\mbox{\boldmath${\rm p}$}}
\newcommand{\veq}{\mbox{\boldmath${\rm q}$}}

\newcommand{\vez}{\mbox{\boldmath${\rm z}$}}

\newcommand{\veL}{\mbox{\boldmath${\rm L}$}}

\newcommand{\veR}{\mbox{\boldmath${\rm R}$}}

\newcommand{\veu}{\mbox{\boldmath${\rm u}$}}

\newcommand{\verho}{\mbox{\boldmath${\rm \rho}$}}

\newcommand{\veta}{\mbox{\boldmath${\rm \eta}$}}
\newcommand{\veB}{\mbox{\boldmath${\rm B}$}}
\newcommand{\veH}{\mbox{\boldmath${\rm H}$}}
\newcommand{\veE}{\mbox{\boldmath${\rm E}$}}

\newcommand{\vegam}{\mbox{\boldmath${\rm \gamma}$}}

\newcommand{\vepi}{\mbox{\boldmath${\rm \pi}$}}

\newcommand{{\Mc}}{\mathcal{M}}
\newcommand{\llan}{\langle\langle}
\newcommand{\rran}{\rangle\rangle}
\newcommand{\lan}{\langle}
\newcommand{\ran}{\rangle}

\begin{document}

\maketitle
\begin{abstract}
The proper-time 4d path integral is used as a starting point to
derive the new explicit parametric form of the  quark-antiquark
Green's function in gluonic and QED fields, entering as a common
Wilson loop. The subsequent vacuum averaging of the latter allows
to derive the instantaneous Hamiltonian. The explicit form and
solutions are given in the case of the $q\bar q$ mesons in
magnetic field.

\end{abstract}

\section{}
 The path-integral formalism in quantum mechanics, created by Feynman
 \cite{1,2} is an important benchmark in the development and in our
 understanding of quantum theory. Many varieties  of this formalism  and new methods to solve the problems, which seemed
 before unsolvable, have been
 suggested since then, see the books  and review papers \cite{3,4,5,6,7,8,9},
 summarizing the modern achievements in this field.

 The extension of the path-integral formalism to the quantum field theory  was done in several directions.
 One  of the most known line of development was   started
 already in \cite{2}, where field variables, i.e. electromagnetic potentials
 $A_i (\vex, t)$, $\varphi (\vex, t)$ play the role of quantum spacial coordinates  $q(t)$, and, the resulting
 path integral is  becoming the functional integral. This line is now a part of
 standard lore, present in  many textbooks, see e.g. \cite{5,6,7,8,9}.

 It is however important, that in all works of this direction the path
 integration concerns spacial coordinates and/or field variables, but not the
 time coordinate, and in this way one can say, that this development is similar
 to the path integrals in quantum mechanics, where time plays the ordering role
 and stays outside of the realm  of fluctuating  variables.

 Another  and more general approach, unifying space and time coordinates in the
 path integral, is  based on the proper time coordinate. The latter was
 introduced  by V.Fock \cite{10} and J.Schwinger \cite{11}, who
 used proper time formalism for the field theory in external electromagnetic
 fields,   however did not exploit path integrals.

 In QED path integrals, based on the   proper time, were suggested in \cite{12} and
 developed  in \cite{13}.

 Path integrals for QCD both in time and space variables, using proper time as
 an ordering variable were suggested in \cite{14,15,16}. The first use of  the QCD
 path integral  for quarks and gluons was done in \cite{17} and exploited  to demonstrate the confinement due to field correlators
 (stochastic confinement) for a review see \cite{18}.

 The full form of the  path integral in QCD for quarks and gluons, based on the
 proper time ordering, was given in \cite{17} for  $T=0$ and in \cite{19} for $T>0$, and different approximations
 were reviewed in \cite{20} for  some relativistic models and in \cite{21} for
 QCD. It was called the Fock-Feynman-Schwinger representation (FFSR) and we
 retain this name in what follows.

 Based on FFSR a new relativistic Hamiltonian was derived in \cite{22, 23} for
 quarks and in \cite{24,25} for gluons, where a new important variable was
 introduced  -- $\omega$, playing the role of the  einbein variable \cite{26}. Its
 average value $\omega_0$ is the average quark (or gluon) energy and explains
 the appearance of the notion of constituent mass in earlier models. The
 relativistic Hamiltonian with einbeins $\omega_i$ allows to calculate all
 low-lying states in QCD: mesons, glueballs, baryons and hybrids from the
 first-principle input: current quark masses, $\alpha_s$ and string tension
 $\sigma$, see \cite{27, 28, 29} for reviews.

 However,  the introduction of $\omega_i$ as einbein variables,  being successful,
 is an approximate procedure,and its limitations and corrections were not
 enough clarified in  the literature. An attempt in this direction was done in
 \cite{30}, where  the fluctuation of the time coordinate in the path integration was substituted
 by the fluctuations integration in $\Delta \omega_i$. The resulting
 expressions for quark decay constants of mesons  in \cite{30} are quite successful
 in comparison with experiment, however the exact scheme of approximations was
 not clearly stated.

An additional impulse for a development in this area was given
recently by the inclusion of high magnetic field $B$ in the
dynamics of QCD and QED, see \cite{31,32,33,34,35} for a recent
papers. In this case $\omega_i$ depend on $B$ and might vanish or
grow fast (depending on quark spin projection), which calls for a
careful analysis of all corrections.

In  the  present paper we undertake such an analysis and rederive
different forms of path integrals and relativistic Hamiltonians
for the QCD and QED systems, typically for quark-antiquark or
atoms, taking into account both QCD and QED dynamics in the first
case.

The first thing we meet confronting $4d$ path integrals, is the
problem of the time-coordinate fluctuation, which necessarily
requires distinguishing average (ordering) time and   fluctuating
time, similar to the old notion of the Zitterbewegung.  We analyze
this phenomenon, comparing the Bethe-Salpeter and path-integral
formalisms and show how the latter can be developed using the
fact, that all dynamics is contained in the Wilson loop formalism
augmented by spin insertions.

\section{Path integral: treating time fluctuations}

We start with the simplest example of a scalar particle in
external field, this problem was considered for QED by Feynman in
\cite{12}.

The  scalar one particle Green's function is (in Euclidean
space-time) \be g(x,y) = \left( \frac{1}{m^2-D^2_\mu}\right)_{xy}
= \int^\infty_0 ds (D^4 z)_{xy} \exp (-K) \Phi (x,y)\label{1}\ee
where $D_\mu = \partial_\mu - ieA_\mu$, \be K= m^2s+ \frac14
\int^s_0 d\tau (\frac{dz_\mu}{d\tau} )^2\label{2}\ee

\be \Phi(x,y) = \exp i e\int^x_y A_\mu dz_\mu,\label{3}\ee and

\be (D^4z)_{xy} \simeq \lim_{N\to \infty} \prod^N_{n=1}\!\int\!\!
\frac{d^4 z(n)}{(4\pi\varepsilon)^2}\! \int\!\!
\frac{d^4p}{(2\pi)^4} e^{ip \left(\sum^N_{n=1}\!
z(n)-(x-y)\right)}, N\varepsilon = s. \label{4} \ee

At this point it is important to stress the difference between the
nonrelativistic quantum-mechanical and relativistic path
integration: in the first case one has $(D^3z) = (D^3z(t))$ in
(\ref{1}) and the time variable $t$ has the ordering character:
the consecutive pieces of trajectory $ \vez (t)$ are ordered by
time. In the relativistic path integral this role is given to the
proper time $\tau, s$ while the ``time'' $z_4(\tau)$ is
fluctuating together with spacial coordinates $\vez(\tau)$. In
terms of any local field theory and Bethe-Salpeter type of
equation this is allowable  and necessary, since any moment of
time $z_4$ appears in the amplitude with a new interaction point,
which may happen before or after the previous interaction point,
thus the points of interaction lie chaotically on   the time axis.
However, from the point of view  of a  stationary process, which
creates the system with a given quantized energy state in the
limit of long time interval, one may think of an averaged
progressive time and averaged trajectories of constituents, where
stochastic time fluctuations are dealt with in a well defined
averaging process. In this way the time-fluctuating relativistic
trajectories are averaged into stationary time-ordered
trajectories, similar to the quantum mechanical ones, where
fluctuations are allowed for spacial coordinates. Correspondingly
one can write \be z_4 (\tau) = \bar z_4 (\tau) + \tilde z_4
(\tau),\label{5}\ee where $\bar z_4 (\tau) \equiv t_E = 2\omega
\tau$ is the averaged time, proportional to the proper time, while
the fluctuating time $\tilde z_4(\tau)$ can be written as a sum of
one-step fluctuations:

\be \tilde z_4 (\tau) = \sum^n_{k=1} \Delta z_4 (k), ~~ \tau =
n\varepsilon, ~~ N\varepsilon =s, ~~ \sum^N_{k=1} \Delta z_4(k)
=0.\label{6}\ee The proper time $s$ is expressed via the total
Euclidean time $T=x_4 - y_4$ and the new variable $\omega$, \be
s=T/2\omega, \label{6a}\ee
and hence the scalar Green's function (\ref{1}) can be rewritten
in the form \be g(x,y) = T \int^\infty_0 \frac{d\omega}{2\omega^2}
D^3z e^{-K(\omega)} \lan \Phi(x,y)\ran_{\Delta z_4}.\label{7}\ee

Here \be K(\omega) = \int^T_0 dt_{E} \left( \frac{\omega}{2} +
\frac{m^2}{2\omega} + \frac{\omega}{2} \left(
\frac{d\vez}{dt_E}\right)^2\right),\label{8}\ee while \be \lan
\Phi (x,y) \ran_{\Delta z_4} = D\Delta z_4 \exp [ ie \int A_i
(\vez(t_E), t_E+ \tilde z_4) dz_i + i e \int A_4 dt_E+ ie \int A_4
d\Delta z_4],\label{9}\ee

\be D\Delta z_4 \equiv \int \frac{dp_4}{2\pi} \prod^n_{k=1} \frac{
d\Delta z_4 (k) }{\sqrt{4\pi \varepsilon}} \exp \left\{
\sum^N_{k=1} \left[ ip_4 \Delta z_4 (k) - \frac14 \frac{(\Delta
z_4 (k))^2}{\varepsilon} + ie \Delta z_4 (k)
A_4\right]\right\}\label{10}\ee

The result of integration in (\ref{10}) can be written as \be \lan
\Phi(x,y) \ran_{\Delta z_4} = \sqrt{ \frac{\omega}{2\pi T} }
\overline{ \Phi (x,y)},\label{12}\ee where $\overline{\Phi} (x,y)$
is the averaged Wilson line, augmented by the fluctuation, \be
\overline{\Phi} (x,y) = \exp \left[ ie \int^{\vex}_{\vey} A_i
(\vez (t_E), t_E) d\vez_i + ie \int^{x_4}_{y_4} A_4 (\vez (t_E),
t_E) dt_E\right]\exp (\Delta S).\label{13}\ee

In the simplest case of the free scalar Green's function $A_\mu
\equiv 0$ and $\overline{\Phi} (x,y) =1$, hence
$$ g_0 (x,y) = \sqrt{\frac{T}{8\pi}}
\int^\infty_0 \frac{d\omega}{\omega \sqrt{\omega}} (D^3
z)_{\vex\vey} e^{-K(\omega)}=$$

\be=\frac{1}{8 \pi^2 T} \int^\infty_0 d\omega \exp \left[ -
\frac{m^2T}{2\omega} - \frac{(\vex-\vey)^2}{2T} \omega -
\frac{\omega T}{2}\right] = \frac{1}{4\pi^2} \frac{m}{u} K_1 (mu),
~~ u^2=T^2+ (\vex-\vey)^2,\label{14}\ee where $K_1 (x)$ is the
Bessel function of the second kind.

 At this point the role of $\omega$ becomes clear, since for large
$T$ the integral in (\ref{14}) can be taken by the stationary
point method with the action \be S(\omega, T) =
\frac{m^2T}{2\omega} + \frac{(\vex-\vey)^2}{2T} \omega +
\frac{\omega T}{2};~~\left. \frac{\partial S (\omega, T)}{\partial
\omega} \right|_{\omega=\omega_0} =0,\label{15}\ee and
$$ \omega_0 = \frac{mT}{\sqrt{(\vex^2-\vey)^2 +T^2}}\to m, ~{\rm for} ~ T\gg
|\vex -\vey|,$$ and  one finally obtains  the standard answer for
large $T$ yielding the asymptotics of the r.h.s. of (\ref{14}) at
large $T$.  \be g_0 (x,y) = \frac{\sqrt{m}}{4\pi^{3/2}
T^{3/2}}\exp (-mT).\label{16}\ee Another form exploits the
Hamiltonian in (\ref{14}), namely, one can use the relation \be
\int (D^3z)_{\vex\vey} e^{-K(\omega)}= \lan \vex |
e^{-H(\omega)T}|\vey\ran\label{17}\ee where \be H(\omega) =
\frac{\vep^2+m^2}{2\omega} + \frac{\omega}{2}.\label{18}\ee

Applying the stationary point method to the integrals (\ref{14}),
(\ref{17}), one obtains at large $T$ the energy eigenvalue \be
\left.\frac{\partial H(\omega)}{\partial
\omega}\right|{\omega=\omega_0} =0;~~ \omega_0 =\sqrt{\vep^2
+m^2}.\label{19}\ee

From (\ref{19}) one can understand, that $\omega$ plays the role
of a virtual particle energy, and the condition (\ref{19}) has the
meaning of the energy shell condition. This interpretation holds
also for the case of $N$  particles with interaction, when the
integrals aver $\prod^N_{i=1} d\omega_i$ are involved. Note, that
in this way $\omega_i$ are not any more approximate einbein
variables, as in our previous works see e.g. \cite{18, 23}.

We now turn to the case of $A_\mu\neq 0$ and remark, that
$A_\mu(\vez, z_4)$ are functions of coordinates, which will be
used later in the process  of vacuum averaging, yielding points of
interaction, correlators etc.,  but at this moment in (\ref{13}),
$\overline{\Phi} (x,y)$ is a set of all possible Wilson lines,
obtained by time fluctuations with the weight, given in
(\ref{10}), see Fig.1 as an illustration.

\begin{figure}
\includegraphics[width= 7cm,height=8cm,keepaspectratio=true]{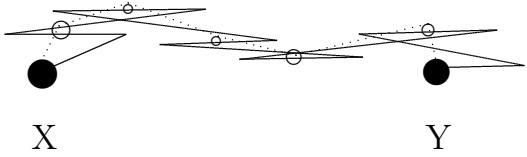}
\caption{The time-fluctuating trajectory in the $(z_1, z_4)$
plane. The points $(z_1 (t_E), t_E)$ are marked by circles and are
connected by average trajectory, depicted by dotted line.}
\vspace{1cm}

\end{figure}

In one particular case, when  $ \vez (t_E)$ is fixed, i.e. the
trajectory is parallel to the $z_4$ axis, all $\Delta z_4$
fluctuations are  washed out, since all  fluctuations cancel each
other, \be \exp (ie \int A_4 (\vez, z_4) d z_4) = \exp (ie \int
A_4 (\vez, t_E) dt_E).\label{20}\ee

The same would happen in the case of QCD, where again overlapping
pieces of Wilson line cancel  each other. We shall come back to
the problem of fluctuating Wilson lines, when we consider gauge
invariant two-body Green's functions.

In the case of the white system of the quark and antiquark of
opposite charges, one must start with the one-body Green's
function

\be S_i (x,y) = (m_i +\hat
\partial - i g \hat A - i e_i \hat A^{(e)})^{-1}_{xy} \equiv(m_i+\hat
D^{(i)})^{-1}_{xy} = (m_1-\hat D^{(i)}) (m^2_i - (\hat
D^{(i)})^2)^{-1}_{xy}.\label{21}\ee

The path-integral representation for $S_i$ is \cite{8} \be S_i
(x,y) = (m_i - \hat D^{(i)})\int^\infty_0 ds_i
(Dz)_{xy}e^{-K_i}\Phi_\sigma^{(i)} (x,y)\equiv (m_i-\hat D^{(i)})
G_i (x,y), \label{22}\ee where \be K_i = m_i^2 s_i +
\frac14\int^{s_i}_0 d\tau_i
\left(\frac{dz_\mu^{(i)}}{d\tau_i}\right)^2,\label{23}\ee
$$\Phi^{(i)}_\sigma
(x,y) =P_AP_F \exp \left( ig \int^x_y A_\mu dz_\mu^{(i)}+ ie_i
\int^x_y A^{(e)}_\mu dz_\mu^{(i)}\right)\times$$ \be \times \exp
\left( \int^{s_i}_0 d\tau_i \sigma_{\mu\nu} (gF_{\mu\nu} + e_i
B_{\mu\nu})\right).\label{24}\ee Here $F_{\mu\nu} $ and
$B_{\mu\nu}$ are correspondingly gluon and c.m. field tensors,
$P_A, P_F$ are ordering operators,  $\sigma_{\mu\nu} =
\frac{1}{4i} (\gamma_\mu\gamma_\nu-\gamma_\nu\gamma_\mu)$. Eqs.
(\ref{21}-\ref{24}) hold for the quark, $i=1$, while  for the
antiquark one should reverse the signs of $e_i$ and $g$. In
explicit form one writes \be \sigma_{\mu\nu} F_{\mu\nu} = \left(
\begin{array}{ll} \vesig\veH&\vesig \veE\\\vesig\veE& \vesig
\veH\end{array}\right),~~ \sigma_{\mu\nu}B_{\mu\nu} = \left(
\begin{array}{ll} \vesig\veB&0\\0&\vesig\veB\end{array}\right).\label{25}\ee

The two-body $q_1 q_2$ Green's function can be written as
\cite{17, 21}

$$ G_{q_1\bar q_2} (x,y) = \int^\infty_0 ds_1 \int^\infty_0 ds_2
(Dz^{(1)})_{xy} (Dz^{(2)})_{xy} \lan  \hat TW_\sigma
(A)\ran_A\times$$ \be \times \exp (ie_1 \int^x_y A^{(e)}_\mu
dz^{(1)}_\mu -ie_2 \int^x_y A^{(e)}_\mu dz^{(2)}_\mu
+e_1\int^{s_1}_0 d\tau_1 (\vesig \veB) -e_2\int^{s_2}_0 d\tau_2
(\vesig \veB)),\label{d26}\ee where \be \hat T = tr (\Gamma_1 (m_1
-\hat D_1) \Gamma_2 (m_2 -\hat D_2)),\label{26a}\ee ``tr'' is the
trace over Dirac and color indices acting on all terms.  Here
$\lan W_\sigma (A)\ran$ is the closed Wilson loop with the spin
insertions and one should have in mind, that color and e.m. spin
insertions in general do not commute, which should be taken into
account when computing spin-dependent part of interaction, see
\cite{36}, in (\ref{d26}) this fact was disregarded.
 \be
W_\sigma (A) = P_a P_F \exp \left[ ig \oint A_\mu dz_\mu + g
\int^{s_1}_0 \sigma^{(1)}_{\mu\nu} F_{\mu\nu} d\tau_1 - g
\int^{s_2}_0 \sigma^{(2)}_{\mu\nu} F_{\mu\nu} d
\tau_2\right].\label{27}\ee

It is important, that the physically meaningful result for the
Green's function is obtained by two different averaging procedures
applied to the total Wilson loop $W=\Phi_\sigma^{(1)} (x,y)
\Phi_\sigma^{(2)} (y,x):$

1) one should overage $W$ over all time fluctuations ;

2) one should average $W$ over all nonperturbative (np) and
perturbative (pert) field configurations with the weight, given by
the standard  QED$+$QCD field actions ,so that the final result is
\be  \llan W \rran \equiv \llan W\ran_{\Delta z_4}
\ran_{A,A^{(e)}}.\label{28}\ee

However, the class of processes  of interest in QCD is very wide,
since any process, starting and finishing with definite hadron
states, such  as formfactors, decays, hadron reactions, needs an
explicit definition of initial and final states as eigenstates of
the Hamiltonian $H(\omega_1, \omega_2)$, and therefore can   use
the formalism, discussed in this paper.

It is clear, that in the fluctuation averaging $\lan W\ran_{\Delta
z_4}$ the result is an  average Wilson loop, passing through the
points $\{ \vez (t_E) + \Delta \vez (t_E), t_E + \Delta t_E\}, ~~
t_E \epsilon (0,T)$, where $\Delta\vez (t_E), \Delta t_E$ depend
on $T, m_1, m_2$ and also on  the concrete field configuration,
which will be averaged in the next averaging process, over vacuum
fields.

One can estimate the average time fluctuation $\Delta t_E$ in the
case of the free relativistic particle propagation.

E.g. assuming the correlation function to have the form

\be f (z_4^{(1)}-z_4^{(2)} )= \exp \left(-\frac{(z_4^{(1)} -
z_4^{(2)})^2}{(\Delta \bar z)^2}\right),\label{29}\ee and taking
into account time fluctuations $z_4^{(1)} = t_E^{(1)} + \tilde z_4
(t_E)$, and integrating over $\Delta  z_4$, one obtains the
increase  of the  correlation time \be (\Delta \bar z)^2 \to
(\Delta z)^2_{\Delta z_4} = (\Delta \bar z)^2 +
\frac{t_E^{(1)}}{2\omega} \sim (\Delta \bar z)^2 +
\frac{T}{2m}.\label{30}\ee

However this result is an artefact of the not accurate definition
of the path-integration measure, when at the ends of the time
interval $\Delta t_E$ the path can  change the direction, implying
infinite  time derivative. Imposing a proper condition on the
magnitude of the derivative,  i.e. with smooth trajectories, the
result would be different. From the point of view of the relation
$\Delta M \Delta t \ga 1,$ one can in principle calculate however
accurate values of masses $M$ for large $T$, and only the coupling
to decay channels, i.e. the width $\Gamma$ should put a lower
limit on the accuracy $\Delta M$.

It is interesting, how this problem occurs in our path-integral
formalism. Indeed, the basic dynamics  which is contained in
$\llan W \rran$, when time fluctuation is supported by
interaction, Eq. (\ref{29}), can be described by the diagram in
Fig.2. Now, from the point of view of Hamiltonian dynamics with
the   trace of the  hypersurface, shown in Fig.2  by a  dotted
line, the Hamiltonian becomes a matrix, with Fock states,
numerating columns and rows,

\be  H_{q\bar q} \to \left( \begin{array} {lll} H_{q\bar q}& \hat V_{12}&... \\
\hat V_{21} & H_{(q\bar q), (q\bar q)_2}&...\\
...&...&...\end{array}\right),\label{31}\ee where nondiagonal
elements are transition operators and diagonal ones define
dynamics (and masses) of more and more complex systems. Therefore,
e.g. $\hat V_{13}$ is responsible for the decay $q\bar q \to
(q\bar q)_1 + (q\bar q)_2$, and hence defines the accuracy of
possible mass determination of the incident state $(q\bar q)$.

\begin{figure}
\includegraphics[width= 7cm,height=8cm,keepaspectratio=true]{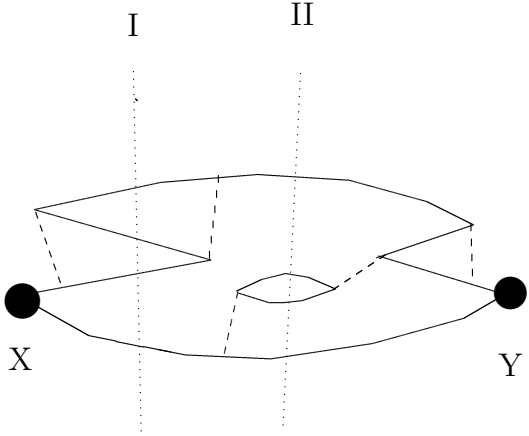}
\caption{The vacuum averaged Wilson lines,  displaying pair
creation in the time fluctuation  process. The hypersurface traces
I and II mark the double quark pair  state of the Hamiltonian}
\vspace{1cm}

\end{figure}

The wave functions  of  (\ref{31}) are actually Fock columns of
different states, e.g.  $\{ \Psi_{q\bar q}, \Psi_{(q\bar q) (q\bar
q),...}\}$, and therefore  the $(q\bar q)$ eigenstates
$\{\Phi_{q\bar q}^{(n)}\}$  are not any more an orthonormal set of
states. As  we shall see, the eigenstates $\Psi_n(\omega_1,
\omega_2)$ will not be orthonormal on the energy shells
$(\omega_1^{(0)}, \omega_2^{(0)})$, different for each $n$. In
this way going from 4d path integral to the relativistic
Hamiltonian 3d formalism one naturally meets the many-channel
Hamiltonian, where diagonal elements correspond to the
fluctuation-averaged trajectories.

\section{From path integral to instantaneous dynamics}

As a result of two averaging processes;  time fluctuation and
vacuum  averaging the basic dynamical input of  the  resulting 3d
path integral -- the doubly renormalized Wilson loop can be
written as   $$ \llan W \rran = Z_W \exp \left\{ - \frac12
\int\int d \pi_{\mu\nu} (1) d\pi_{\lambda\sigma} (2) \left [ g^2
\lan F_{\mu\nu} (1) F_{\lambda \sigma} (2)\ran +\right. \right.$$
\be \left.\left.e^2 \lan F^{(e)}_{\mu\nu} (1) F_{\lambda
\sigma}^{(e)} (2) \ran \right ]+ O(FFF)\right\}\label{32}\ee where
$$ d\pi_{\mu\nu} \equiv d s_{\mu\nu} + \sigma^{(1)}_{\mu\nu}
\frac{dt^{(1)}_E}{2\omega_1} - \sigma^{(2)}_{\mu\nu}
\frac{dt^{(2)}_E}{2\omega_2},$$ and the integration
$ds_{\mu\nu}$is done over the minimal area $S_{\min}$ inside the
time-averaged trajectories of quark and antiquark $\bar L_1$ and
$\bar L_2$. Note, that in addition to the time-fluctuation
smearing discussed above, there is also nonperturbative smearing,
provided by the np field correlators.

Indeed, the quadratic (Gaussian) color field correlators can be
written as \cite{19}

 $$ \frac{g^2}{N_c}\lan\lan {\rm Tr} E_i(x)\Phi
E_j(y)\Phi^\dagger\ran\ran=\delta_{ij}\left(D^E(u)+D_1^E(u)+u^2_4\frac{\partial
D_1^E}{\partial u^2}\right)+ u_iu_j\frac{\partial D_1^E}{\partial
u^2},$$
$$\frac{g^2}{N_c}\lan\lan {\rm Tr} H_i(x)\Phi
H_j(y)\Phi^\dagger\ran\ran=\delta_{ij}\left(D^H(u)+D_1^H(u)+\veu^2\frac{\partial
D_1^H}{\partial\veu^2}\right)- u_iu_j\frac{\partial
D_1^H}{\partial u^2},\label{Hs0}$$ \be\frac{g^2}{N_c}\lan\lan {\rm
Tr} H_i(x)\Phi E_j(y)\Phi^\dagger\ran\ran=\varepsilon_{ijk}
u_4u_k\frac{\partial D_1^{EH}}{\partial u^2}, \label{33} \ee where
$D^E, D^H$ are purely np correlators, and $D^{E,H}_1$ contain
perturbative part.  The same type of equations, but with
replacement $\frac{g^2}{N_c} \to e^2$
 and keeping only $D^{E,H}_1$ holds also for e.m. correlators.
  Note, that at
 zero temperature colorelectric and colormagnetic correlators coincide, note
 also that np correlators $D^E, D^H$ are due to Euclidean vacuum fields.

 The explicit form of  perturbative correlators $D_1^{E,H}$  to lowest order in $\alpha_s$ is
 \be D_1^E (x) = D_1^H(x) = \frac{16\alpha_s}{3\pi x^4} +
 O(\alpha_s^2),\label{34}\ee
 while for e.m. fields one should replace $\frac43 \alpha_s \to \alpha$.

At this point it is important to realize, that the correlators
depend on space and time intervals, e.g. $D(\vez^{(1)} -
\vez^{(2)}, t_E^{(1)} - t_E^{(2)})$ and $\llan W \rran $ in  Eq.
(\ref{33}) even  after fluctuation averaging implies nonlocal in
time dynamics, e.g. the term $\int \int e^z \lan F^{(e)}_{\mu\nu}
(1) F^{(e)}_{\lambda\sigma} (2) \ran ds_{\mu\nu} (1)
ds_{\lambda\sigma} (2)$ stands actually for a photon exchange
diagram. We are now going to replace this time nonlocal
interaction by the instantaneous one, which is easily done in the
correlator language, simply by integrating in (\ref{33}) all
correlators over time differences, $t^{(1)}_E - t^{(2)}_E$,
$$ dt_E^{(1)} dt_E^{(2)}=dt_E d (t_E^{(1)} - t_E^{(2)}),  dt_E = d\frac{t_E^{(1)}
+t_E^{(2)}}{2}.$$

It is important, that the main part of our interaction, the
confining interaction, is ensured by the  correlator
$D^E(\sqrt{t^2+ \ver^2})$, which has a very small correlation
length $\lambda$, as was shown on the lattice \cite{36} and
analytically \cite{37}. $ D^E(t) \sim e^{-t/\lambda}, t\ga
\lambda, ~~ \lambda \sim 0.1 $ fm, and therefore the transition to
the instantaneous dynamics  is done on small averaging interval
$\Delta t \sim \lambda$. Therefore for all processes with momentum
(energy) transfer $\Delta Q$, satisfying $\Delta Q \lambda \la 1$,
this transition of np confining mechanism to the instantaneous
dynamics is allowable. The case of gluon exchange is similar to
the Coulomb interaction, where the instantaneous approximation in
the Bethe-Salpeter equation is known as the Salpeter equation and
is widely used in the literature. We shall mostly use the
one-gluon exchange (OGE) interaction as a perturbation and
therefore our transition to the instantaneous dynamics is
justified.

For the case of the zero angular momentum (see in \cite{23} the
general derivation) one can write for the instantaneous straight
line  $w_\mu (t,\beta) = z_\mu^{(1)} (t) \beta = z_\mu^{(2)} (t)
(1-\beta) , ~ 0 \leq \beta \leq 1,$ and e.g. $ds_{\mu 4} =
(z_\mu^{(1)} (t) - z_\mu^{(2)} (t)) d\beta d t$.

 For  zero angular momentum one can simplify the integration over the area of
 the minimal surface in (\ref{32}) and obtain the result, neglecting
 spin-containing terms in (\ref{32}) for  the moment,
 \be \llan W \rran = Z_W \exp (-\int^T_0 [V_0(r(t_E))] )
 dt_E),\label{35}\ee
 where $r(t_E) = |\vez_1 (t_E) - \vez_2(t_E)|$, and

\be V_0 (r) = V_{conf} (r) + V_{OGE} (r), \label{36}\ee \be
V_{conf} (r) = 2r \int^r_0 d\lambda \int^\infty_0  d\nu D(\lambda,
\nu) \to \sigma r, (r\to \infty),\label{37}\ee

\be \sigma=2\int_0^\infty d\nu\int_0^\infty d\lambda
D(\nu,\lambda), \label{38} \ee

 \be V_{OGE} = \int^r_0 \lambda d\lambda \int^\infty_0
d\nu D_1^{ pert} (\lambda, \nu) =- \frac43
\frac{\alpha_s}{r}\label{39}\ee

As a result one can write for the product of $q\bar q$ Green's
functions (we omit renormalization $Z$ factors, Fock amplitude
coefficients, and ordering operators for simplicity)

\be \left( \frac{1}{(m^2_1 - \hat D^2_1) (m^2_2 - \hat
D^2_2)}\right)_{xy} = \frac{T}{8\pi} \int^\infty_0
\frac{d\omega_1}{\omega_1^{3/2}} \int^\infty_0
\frac{d\omega_2}{\omega_2^{3/2}} (D^3 z_1)_{\vex\vey} (D^3
z_2)_{\vex\vey} e^{- A(\omega_1, \omega_2, \vez_1,
\vez_2)},\label{40}\ee where $A\equiv K_1 (\omega_1)  + K_2
(\omega_2) + \int V_0 (r(t_E)) dt_E$, and $$ K_i(\omega_i) =
\frac{m_i^2 +\omega_i^2}{2\omega_i} T + \int^T_0 dt_E
\frac{\omega_i}{2} \left( \frac{d\vez^{(i)}}{dt_E}\right)^2$$ We
can also introduce here the two-body 3d Hamiltonian $H(\omega_1,
\omega_2, \vep_1, \vep_2$) and rewrite (\ref{40}) as

\be \left( \frac{1}{(m^2_1 - \hat D^2_1) (m^2_2 - \hat
D^2_2)}\right)_{xy} = \frac{T}{8\pi} \int^\infty_0
\frac{d\omega_1}{\omega_1^{3/2}} \int^\infty_0
\frac{d\omega_2}{\omega_2^{3/2}} \lan \vex |e^{- H(\omega_1,
\omega_2, \vep_1, \vep_2)T}|_{\vey}\ran.\label{41}\ee where $H$ is
obtained in a standard way from the action $A(\omega_1,\omega_2,
\vez_1, \vez_2)$ (we omit all e.m. fields except for external
magnetic fields $\veB$) \be H= \sum^2_{i=1} \frac{(\vep^{(i)} -
\frac{e_i}{2} (\veB \times \vez^{(i)}))^2 + m^2_i + \omega^2_i -
e_i \vesig^i \veB}{2\omega_i} +V_{0} (r) + V_{ss}  + \Delta M_{SE}
\label{42}\ee and $V_0$ is given in (\ref{36}). The spin-dependent
part of $H, V_{ss}$ is obtained perturbatively from
$\sigma_{\mu\nu} F_{\mu\nu}$ terms in (\ref{27}), and is
calculated in the presence of m.f. in \cite{36}. It is considered
as a perturbative correction and is a relativistic generalization
of the standard hyperfine interaction, $$V_{ss} ( r) =
\frac{1}{4\omega_1\omega_2} \int \lan \sigma_{\mu\nu}^{(i)}
F_{\mu\nu}(x) \sigma_{\rho\lambda}^{(2)} F_{\rho\lambda}(y)\ran
d(x_4-y_4).$$ Its explicit form is given in \cite{38}. Finally,
the correction $\frac{\lan \sigma^{(i)} F(x) \sigma^{(i)}
F(y)\ran}{4\omega_1\omega_2}$, where $i$ refers to the same quark
(antiquark) yields the spin-independent self-energy correction
$\Delta M_{SE}$ which
 was calculated
earlier \cite{39}  and for zero mass quarks and no m.f. is  \be
\Delta M_{SE} =- \frac{3\sigma}{2\pi\omega_1}-
\frac{3\sigma}{2\pi\omega_2}.\label{44}\ee For the case of nonzero
m.f. the resulting $\Delta M_{SE}$ is given in \cite{38}. We can
now write the total Green's function of $q_1\bar q_2$ system,
denoting by $Y$ the product of projection operators $Y= \Gamma
(m_1 - \hat D_1) \Gamma (m_2 - \hat D_2)$, \be m_1 - \hat D_1 =
m_1 - i\hat p_1=m_1 +\omega_1 \gamma_4 - i \vep\vegam ,~~ m_2 -
\hat D_2= m_2 - \omega_2 \gamma_4 - i\vep\vegam, \label{45}\ee
where $\vep$ is the quark 3 momentum in the c.m. system.

As a result one has$$\int d^3 (\vex-\vey) G(x,y) = \int d^3
(\vex-\vey)
 tr \left( \frac{4Y_\Gamma}{(m^2_1 - \hat D^2_1) (m^2_2 - \hat
D^2_2)}\right)_{xy}=$$ \be = \frac{T}{ 2\pi} \int^\infty_0
\frac{d\omega_1}{\omega_1^{3/2}} \int^\infty_0
\frac{d\omega_2}{\omega_2^{3/2}}
 \lan Y_\Gamma\ran  \lan \vex |e^{- H(\omega_1, \omega_2, \vep_1,
\vep_2)T}|_{\vey}\ran,\label{46}\ee

We have used  in (\ref{46}) the relations  $ 4\lan Y\ran = tr \lan
\Gamma (m_1 - i\hat p_1) \Gamma(m_2 - i\hat p_2))$, and neglect
spin dependent terms in $H$; we have taken into account, that
$D_\mu$ acting on Wilson line, i.e. $D_\mu$ $\exp (ig \int^x A_\mu
dz_\mu)\Lambda$ yields $\exp (ig \int^x A_\mu dz_\mu)
\partial_\mu \Lambda$. The c.m. projection of the Green's function yields  \be
\int d^3 (\vex-\vey) \lan \vex| e^{-H(\omega_1, \omega_2, \vep_1,
\vep_2)T}|_{\vey}\ran = \sum_n \varphi^2_n (0) e^{- M_n (\omega_1,
\omega_2) T},\label{48}\ee  see Appendix 1 for explicit separation
of relative coordinates, Eq. (\ref{a1.8})--(\ref{a1.11}). Here
$M_n (\omega_1, \omega_2)$ is the eigenvalue of $H(\omega_1,
\omega_2, \vep_1, \vep_2)$ in the c.m. system, where $\veP =
\vep_1 + \vep_2 =0; ~~ \vep_1 =\vep=-\vep_2$.

The integrals over $d\omega_1, d\omega_2$   for $T\to \infty$  can
be performed by the stationary point method, namely  one has $$
\int G(x,y) d^3 (\vex-\vey) = \frac{T}{2\pi} \int^\infty_0
\frac{d\omega_1}{\omega_1^{3/2}}\int^\infty_0
\frac{d\omega_2}{\omega_2^{3/2}} \sum_ne^{-M_n (\omega_1,
\omega_2)T}\varphi_n^2(0) \lan Y \ran $$

\be = \sum_n \frac{  e^{-M_n (\omega_1^{(0)}, \omega_2^{(0)}
)T}\varphi_n^2(0)\lan Y\ran }{  \omega_1^{(0)} \omega_2^{(0)}
\sqrt{ (\omega_1^{(0)} M^{"}_n(1))(\omega_2^{(0)} M^{"}_n(2))}},
\label{49}\ee where
  \be \left.\frac{\partial M_n (\omega_1, \omega_2)}{\partial \omega_i}
\right|_{\omega_i = \omega_i^{(0)}}=0, ~~ \left. M_n^{"} (i)
=\frac{\partial M_n (\omega_1, \omega_2)}{\partial \omega_i^2}
\right|_{\omega_i = \omega_i^{(0)}},\label{50}\ee
and we have neglected the mixed terms
$\frac{\partial^2M_n}{\partial \omega_1
\partial \omega_2}$ for simplicity, however should keep them in concrete
calculations: see exact result in Appendix 1. Comparing the
results (\ref{48}), (\ref{49}) with the definitions of quark decay
constants $f^n_\Gamma$,
\begin{eqnarray}
 \int G_\Gamma(x) d^3 \vex & = &
 \sum_n \int d^3 \vex \lan 0 | j_\Gamma| n\ran \lan n |j_\Gamma|0\ran
 e^{i\veP    \vex -M_nT}\frac{d^3\veP  }{2M_n(2\pi)^3}
\nonumber \\
 & = & \sum_n \varepsilon_\Gamma \otimes \varepsilon_\Gamma
 \frac{(M_n f_\Gamma^n)^2}{2M_{n}} e^{-M_nT},
\label{51}
\end{eqnarray}
where for   $\Gamma =\gamma_\mu, ~ \gamma_\mu\gamma_5$ \be
 \sum_{k=1,2,3} \varepsilon_\mu^{(k)} (q) \varepsilon_\nu^{(k)} (q) =
 \delta_{\mu\nu}-\frac{q_\mu q_\nu}{q^2},
\label{52} \ee
and $\varepsilon_\Gamma =1$ for $\Gamma=1,\gamma_5$, one obtains
the expression for $f^n_\Gamma$ (to lowest order in $V_{ss}$)

\be (f_\Gamma^n)^2=\frac{ N_c \lan Y_\Gamma\ran | \varphi_n
(0)|^2}{ \omega_1^{(0)} \omega_2^{(0)} M_n \xi_n}, ~~ \xi _n
\equiv \sqrt{(\omega_1^{(0)}M^{"}_n(1))(\omega_2^{(0)}
M^{"}_n(2))},
 \label{53a} \ee

 It is interesting, that numerical estimates using (\ref{53a}) and  (\ref{a1.18})  are close to
 those, obtained in \cite{30}.

\section{Relativistic Hamiltonians of a meson in magnetic field}

The resulting relativistic Hamiltonian in the instantaneous limit
ia given in (\ref{42}) and can be written as \be H-\sum^2_{i=1}
\frac{(\vep^{(i)} -\frac{e_i}{2} (\veB\times \vez^{(i)}))^2+ m^2_i
+\omega^2_i - e_i \vesig^{(i)} B}{2\omega_i} + U (\vez^{(1)} -
\vez^{(2)}, ~~
\vesig^{(1)},~\vesig^{(2)},\omega_1,\omega_2)\label{53}\ee where
\be U=V_0(r) +V_{ss} +\Delta M_{SE}\label{54}\ee

We shall be interested  in the spectrum of the $q_1\bar q_2$
system in the magnetic field $\veB$, but
 before that we shall test the general form of the Hamiltonian $H(\omega_1,\omega_2, \vep_1\vep_2)$
and its eigenvalues, obtained at the stationary point values
$\omega_1^{(0)},\omega_2^{(0)}$.

We start with the case of $ \veB =0$ and
$U=-\frac{Z\alpha}{|\vez^{(1)}-\vez^{(2)}|}.$ Separating the total
and relative momenta and coordinates,

\be \veR = \frac{\omega_1 \vez^{(1)}+
\omega_2\vez^{(2)}}{\omega_1+\omega_2}, ~~ \veta
=\vez^{(1)}-\vez^{(2)};~~\vepi=\frac{1}{i}\frac{\partial}{\partial
\veta},\label{55}\ee and $\veP=\vep^{(1)}+\vep^{(2)}$, one obtains
in () with $\veB=0$ \be
H=\frac{\veP^2}{2(\omega_1+\omega_2)}+\frac{\vepi^2}{2\tilde
\omega} + U (\veta)+ \sum_{i=1,2}
\frac{m^2_i+\omega^2_i}{2\omega_i}.\label{56}\ee

\begin{enumerate}
\item  The first example is the relativistic electron with mass
$m_1$ in the Coulomb field of a
 heavy  atom of mass $m_2$ with charge $Ze$,
$ U(\eta) = -\frac{Z \alpha}{\eta}$. For $\veP=0$ one has for the
ground state

\be M(\omega_1,\omega_2) =\sum_{i=1,2}
\frac{m^2_i+\omega^2_i}{2\omega_i}-\frac{\tilde
\omega(Z\alpha)^2}{2}.\label{57}\ee

Minimizing in $\omega_1$ for $m_2\gg m_1$ one obtains \be M\approx
m_2 + m_1 \sqrt{1-(Z\alpha)^2},\label{58}\ee which coincides with
the exact answer from the Dirac equation.

\item As a second example we consider electron-positron system,
then from the same Hamiltonian (\ref{56})
 for $\veP=0$ and $m_1=m_2=m$ one obtains after minimization
\be M= 2 m \sqrt{1-\frac{\alpha^2}{4}}\approx 2m -
\frac{m\alpha^2}{4},\label{59}\ee which looks correct, at least
in the expansion in $\alpha$.

\item In the next example we consider the noninteracting $q_1\bar
q_2$ system in constant
 magnetic field $\veB$ along the  $z$ axis. For $U=0$ one can solve one-body problem  for each quark in m.f. with the
result for the lowest Landau levels (LLL)

\be M(\omega_1,\omega_2)= \sum_i\frac{m^2_i +\omega^2_i + eB
(2n_i+1)- e_i\vesig^{(i)}\veB+
(\vep^{(i)}_{z})^2}{2\omega_i}\label{60}\ee and after minimization
one has

\be M(\omega_1^{(0)},\omega_2^{(0)})= \sum_i\sqrt{
(\vep^{(i)}_{z})^2+m^2_i + eB (2n_i+1)-
e_i\vesig^{(i)}\veB}\label{61}\ee

\end{enumerate}

We turn now to the general case of the $q_1\bar q_2$ system and
consider first the  case of a neutral system, $e_1=-e_2=e$. In
terms of total and relative momenta the Hamiltonian has the form

\be H_{q_1q_2} = H_B+H_\sigma +U\label{62}\ee

$$ H_B= \frac{1}{2\omega_1}\left[ \frac{\tilde \omega}{\omega_2} \veP+\vepi -\frac{e_1}{2} \veB \times (\veR+
\frac{\tilde \omega}{\omega_1} \veta)\right]^2+$$

\be + \frac{1}{2\omega_2}\left[ \frac{\tilde \omega}{\omega_1}
\veP-\vepi -\frac{e_2}{2} \veB \times (\veR- \frac{\tilde
\omega}{\omega_2} \veta)\right]^2\label{63}\ee

\be H_\sigma= \sum_{i=1,2}\frac{m^2_i +\omega^2_i-
e_i\vesig^{(i)}\veB}{2\omega_i}\label{64}\ee

The $\veR$ dependence  for (\ref{62}) in the case, when $e_1=-e_2$
can be factorized out in the way, discovered long ago \cite{40}

\be \Psi(\veR,\veta) = \varphi(\veta) \exp (i\veP
\veR-\frac{ie}{2} (\veB\times \veta) \veR) \label{65}\ee and for
$\varphi(\veta)$ one obtains the equation: \be ( H_0 + H_\sigma+
U) \varphi(\eta) = M\varphi(\eta),\label{66}\ee

where $H_0$ is \be H_0 =\frac{1}{2\tilde \omega} \left(
-\frac{\partial^2}{\partial\veta^2}+ \frac{e^2}{4}(\veB\times
\veta)^2\right).\label{67}\ee

In this way $H_0$ adds to the confining well with OGE and other
terms also an oscillator potential.

 One can replace for simplicity the linear confining term by the oscillator
 potential, $V_{conf} = \sigma \eta \to \tilde V_{conf} \equiv \frac{\sigma}{2}
 \left( \frac{\eta^2}{\gamma} + \gamma\right)$, where $\gamma$ satisfies
 stationary point condition $\frac{\partial M}{\partial \gamma} |_{\gamma=
 \gamma_0} =0$,  which ensures some 5\% accuracy of this replacement. Then the
 lowest eigevalue $\bar M$ of the basic part of Hamiltonian, $\bar H = H_0 +
 H_\sigma + \tilde V_{conf}$, is \be \bar M (\omega_1,\omega_2, \gamma) =
 \varepsilon_{n_\bot, n_z} + \sum_{i=12,} \frac{m_i^2 +\omega_i^2 - e_i
 \vesig^{(i)} \veB}{2\omega_i},\label{70a}\ee where $e_1=e=-e_2$, and

\be \varepsilon_{n_\bot, n_z} = \frac{1}{2\tilde \omega} \left[
\sqrt{ e^2 B^2 + \frac{4\sigma\tilde \omega}{\gamma}} (2n_\bot
+1)+ \sqrt{\frac{4\sigma \tilde \omega}{\gamma}}\left(n_z +
\frac12\right)\right] + \frac{\gamma \sigma}{2}, \label{71a}\ee

We turn now to the case of charged two-body system in m.f.,  and
here one can consider
 two different situation. In the first case, when $e_1=e_2= {e} $ and also $m_1=m_2$ (and hence $\omega_1^{(0)}=
\omega_2^{(0)}$) one can do an exact factorization of $\veR$ and
$\veta$.

\be H_B =\frac{P^2}{2(\omega_1 +\omega_2)} - \frac{e\veP (\veB
\times \veR)}{\omega_1 +\omega_2} + \frac{e^2}{8\tilde \omega}
(\veB\times \veR)^2 + \frac{\pi^2}{2\tilde \omega} + \frac{e^2
(\veB \times \veta)^2 ( \omega^3_1 + \omega^3_2)}{8 (\omega_1 +
\omega_2)^2 \omega_1 \omega_2} + \Delta H_B (\omega_1,
\omega_2);\label{68}\ee

$H_\sigma $  is given in (\ref{64}), and  $\Delta H_B$ is

$$ \Delta H_B (\omega_1, \omega_2) = - \frac{\omega^2_2-\omega^2_1}{\omega_1
\omega_2 (\omega_1 + \omega_2)} \frac{e}{2} \vepi (\veB\times
\veta) - \frac{\omega_2 -\omega_1}{\omega_1\omega_2} \frac{e}{2}
\vepi (\veB\times \veR)-$$

\be - \frac{\omega_2-\omega_1}{ (\omega_1 \omega_2 )} \frac{e}{2}
\veP (\veB\times \veta) +\frac{(\omega_2^2
-\omega_1^2)}{(\omega_1+\omega_2)^2\omega_1\omega_2} \frac{e^2}{4}
(\veB\times \veR) (\veB\times \veta)\label{69}\ee

 For $\omega_1 = \omega_2, \Delta H_B$ vanishes and the Hamiltonian has the
 form

$$ H= \frac{P^2}{4\omega} - \frac{e(\veP(\veB \times \veR))}{2\omega} +
\frac{e^2}{4\omega} (\veB\times \veR)^2 + \frac{\pi^2}{\omega} +
\frac{e^2}{16 \omega} (\veB\times \veta)^2+$$\be +  \frac{2m^2 + 2
\omega^2 - e (\vesig_1 + \vesig_2) \veB}{2\omega} +
\frac{\sigma}{2} \left(\frac{\eta^2}{\gamma} + \gamma\right) +
V_{\rm OGE} + V_{ss} + \Delta M_{SE}.\label{70}\ee

The lowest eigenvalues of the Hamiltonian
 (\ref{70} ) are

$$  M= \frac{m^2+\omega^2}{\omega} + \lan V_{\rm OGE} \ran + \lan V_{ss} \ran
+ \lan \Delta M_{SE}\ran+$$ \be + \frac{eB}{2\omega} (2 N_\bot +1)
+ \sqrt{\left( \frac{eB}{2\omega}\right)^2 +
\frac{2\sigma}{\gamma_0\omega}} (2n_\bot +1) + (n_\parallel +
\frac12) \sqrt{\frac{2\sigma}{\gamma_0 \omega}}-
\frac{e(\vesig_1+\vesig_2)\veB}{2\omega} +
\frac{\gamma_0\sigma}{2}, \label{72}\ee

We now turn to the general case of a charged $q_1 \bar q_2$
system, when $e_1\neq e_2$, and write the full instantaneous
Hamiltonian as in (\ref{62}) -(\ref{64}), but with arbitrary $e_1$
and $e_2, e_1+e_2=e$ and $e$ is the total charge of the meson.

In this case the simple factorization form (\ref{65}) does  not
work, and one must instead to make a first step towards
factorization, namely one must associate the c.m. motion in m.f.
with the total charge $e$ of the system. This is done in the
following form, discussed previously  in \cite{33}

\be \Psi(\veta,\veR)=\exp(i\Gamma) \varphi (\veta,
\veR),\label{73}\ee

\be \Gamma = \mathbf{P}\mathbf{R} - \frac{\bar e}{2} (\veB \times
\veta) \veR, ~~ \bar e = \frac{e_1-e_2}{2} \label{74}\ee
 and the  resulting
  Hamiltonian from the relation    $H_0\Psi=\exp
(i\Gamma) H'_0\varphi$, is

$$ H'_0=\frac{\veP^2}{2(\omega_1+\omega_2)}+\frac{(\omega_1+\omega_2)\Omega^2_R\veR^2_\bot}{2}
+\frac{\vepi^2}{2\tilde \omega}+ \frac{\tilde \omega
\Omega^2_\eta\veta^2_\bot}{2}+X_{LP}\veB\veL_P+$$

$$+X_{L_{\eta}}\veB\veL_\eta+X_1\veP(\veB\times\veta)+X_2(\veB\times \veR)\cdot(\veB\times\veta)+ $$

\be+ X_3\vepi(\veB\times \veR)+\frac{m^2_1+\omega^2_1}{2\omega_1}+
\frac{m^2_2+\omega^2_2}{2\omega_2}.\label{75}\ee


\be\Omega^2_R=
B^2\frac{(e_1+e_2)^2}{16\omega_1\omega_2}\label{76}\ee \be
\Omega^2_\eta=\frac{B^2}{2\tilde \omega(\omega_1+\omega_2)^2}
\left[ \frac{(e_1\omega_2+\bar
e\omega_1)^2}{2\omega_1}+\frac{(e_2\omega_1-\bar e\omega_2)^2}
{2\omega_2}\right].\label{77}\ee  Here  all coefficients $X_i
(i=1,2,3)$   given  explicitly in the Appendix 2 of \cite{33}.

Treating the    terms $X_1, X_2, X_3$  as a perturbation  $\Delta
M_X$,

  \be \Delta M_X = \lan X_1 \veP
(\veB\times \veta) + X_2 (\veB \times \veR) (\veB \times \veta) +
X_3 \vepi (\veB \times \veR) \ran,\label{78}\ee one can write the
total energy eigenvalues    $M_n^{(0)}$ of the  Hamiltonian
  $\bar H$ in (\ref{70a}) as \be M_n^{(0)} = M^{(0)} (\veP) + M^{(0)} (\vepi)+
  \Delta M_X +   H_\sigma  \label{79}\ee
  where
  \be M^{(0)} (\veP) = \frac{P^2_z}{2(\omega_1 +\omega_2)} + \Omega_R (2
  n_{R_\bot}+1) +X_{LP} \veL_P \veB,\label{80}\ee
  $M^{(0)}(\vepi)$ is the eigenvalue of the operator $H_\pi$,
  \be H_\pi = \frac{\vepi^2}{2\tilde\omega}+ \frac{\tilde \omega
  \Omega^2_\eta \veta^2_\bot}{2} +X_{L_\eta} \veB \veL_\eta + V_{conf}
  +V_{OGE}.\label{81}\ee

We have written above the most general   forms of instantaneous
Hamiltonians in the  external m.f.  It is  seen, that to a good
accuracy the dynamical contributions of e.m. and color fields can
be separated, except in the OGE and spin-dependent terms,  and as
shown in  \cite{38}, the m.f. contribution to the both terms is
decisive at large $eB$.

\section{Discussion of results}

We have started with the general 4d proper-time path integral for
the Green's function of  a quark  and an antiquark in gluonic
$(A_\mu, F_{\mu\nu}) $ and e.m. $(A_\mu^e, B_{\mu\nu}$) fields.
These fields are contained  in the generalized Wilson loop $W$
with inclusion of spin-field operators  $(
\sigma_{\mu\nu}(F_{\mu\nu} + B_{\mu\nu}))$.

After vacuum averaging procedure in the partition function, the
averaged Wilson loop  $\lan W\ran_{A, A^{(e)}}$ contains all
possible interactions, including internal quark loops from the
terms $tr \ln (m^2_i - \hat D^2_i)$ in the partition function.

As a first step we have traded the particle proper times for the
Euclidean (ordering) times $t_E^{(1)}, t_E^{(2)}$ and performed
path integration over fourth particle coordinates $z_4, \bar z_4$,
which is  physically the time fluctuations around  $t_E^{(1)},
t_E^{(2)}$. We have shown, that this time-fluctuation integration
leads to the 3d path integrals with the action (or Hamiltonian in
the Hamiltonian form of path integral) which is a matrix  in the
Fock states. The resulting 3d path integrals   are  integrals
over new parameters $\omega_1, \omega_2$, and the spectrum of the
$q_1\bar q_2$ system can be found for large times by a stationary
point  procedure in $\omega_1, \omega_2$.

In this way one  is going from the 4d formalism to the
multichannel 3d formalism with   an additional $\omega$-
integration for each  particle.

As a next step we have observed that the interaction appearing in
the averaged Wilson loop, $\lan W \ran_{A,A^{(e)}}$, has the form
of field correlators $\lan F_{\mu\nu} (x) F_{\lambda\sigma} (y)
\ran, $ $ \lan B_{\mu\nu} (x) B_{\lambda \sigma} (y)\ran$, and the
first correlator has a very small correlation length $\lambda\sim
0.1$ fm (found on the lattice \cite{36}  and in analytic
calculations  \cite{37}). This allows  to go over to the
instantaneous dynamics, when the bolocal or multilocal)
interaction $\lan F (x) F(y)\ran$ is replaced by the time-averaged
potentials $V(\vex-\vey) = \int d (x_4 - y_4) \lan F (x) F(y)\ran$
, and this is valid when the basic parameter, defining the quark
trajectory, string tension $\sigma$ satisfies $\sigma \lambda^2\ll
1$, so that typical time length on trajectory $t_0 \sim
\frac{1}{\sqrt{\sigma}}$ is much larger than $\lambda$. Note, that
this condition is opposite to the one, used for validity of the
OPE and QCD sum rules.

As a result one obtains the instantaneous relativistic Hamiltonian
$H(\omega_1, \omega_2)$ depending on two parameter $\omega_1,
\omega_2$ (for the $(q_1 \bar q_2)$ Hamiltonian matrix element)
and the actual spectrum is obtained from the eigenvalues
$M_n(\omega_1, \omega_2)$ at the stationary points
$\omega_1^{(0)}, \omega_2^{(0)}$. Note, that these points are
different for different $n=0,1,2...$

We have checked the  results in section 4 for several simple
systems and found good agreement with known results. Moreover,
this formalism for eigenvalues was being used for more than 20
years in many papers, a small part of which was cited here, and
the results  in all systems, mesons, baryons, hybrids and
glueballs are well  compared with experimental and lattice ones.

The important new element in this paper is the rigorous derivation
of the integral representation for the $(q_1\bar q_2)$ Green's
function Eqs. (\ref{40}), (\ref{41}), (\ref{49}), which gives  a
new meaning to the parameters $\omega_1, \omega_2$, and allows not
only calculate spectrum, but also the Green's function itself.

 As an important application  of the developed formalism, we have derived in
 section 4 the explicit form of Hamiltonians of the $(q_1 \bar q_2)$ system in
 the constant m.f. $\veB$, and defined the main part of the spectrum for
 neutral and charged mesons.

 These results have been used for the explicit numerical evaluation of the
 $\rho$-meson spectra in \cite{41} , which are in reasonable agreement with existing
 lattice data. Moreover, the same formalism was extensively exploited in
 \cite{35}
 for calculation of chiral condensate, and in \cite{33} for magnetic moments.

 Actually, the field of possible applications of our method in QCD and QED is
 enormous, and  the  method is especially simple in the cases, when only spectral properties
 are of interest. This is clearly seen, when one compares this method with the
 Bethe-Salpeter equation. In the last case one is facing the  problems of the
 relative time and insufficiency of the ladder kernel already in the QED case.

 In the QCD case the use of the Bethe-Salpeter equation  is in addition
 associated with
 the vector propagator form of confinement, which is physically not consistent,
 or  with some phenomenological form, and in this way the method loses its
 fundamental character. On the contrary, the very short-correlation property of
 confinement suits perfectly to establish the validity of instantaneous
 Hamiltonian formalism and allows for  an accurate and simple procedure.

The author  is grateful to M.A.Andreichikov, A.M.Badalian,
S.I.Godunov, B.O.Kerbikov, V.D.Orlovsky, A.E.Shabad,
M.I.Vy\-sotsky for many useful discussions.

\vspace{2cm}
 \setcounter{equation}{0}
\renewcommand{\theequation}{A \arabic{equation}}

\hfill {\it  Appendix  1.}

\centerline{\bf \large Derivation of the general expression for
the $q_1 \bar q_2$ Green's function}

 \vspace{1cm}

\setcounter{equation}{0} \def\theequation{A.1 \arabic{equation}}

 We start with
the general definition for the $q_1 \bar q_2$ Green's function in
the vacuum gluonic  and external e.m. fields $$ G(x,y) = \lan tr
\Gamma S_1 (x,y) \bar \Gamma \bar S_2 (y,x)\ran_A=$$ \be = \lan tr
\Gamma \frac{(m_1 - \hat D_1)} {m^2_1 - \hat D^2_1} \bar
\Gamma\frac{(m_2 -\hat{\bar D}_2)}{(m_2^2- \hat{\bar
D}^2_2)}\ran_A=\label{a1.1}\ee \be = 4 \int^\infty_0 ds_1
\int^\infty_0 ds_2 (D^4 z^{(1)} D^4z^{(2)})_{xy} e^{-K_1 -K_2}
\lan YW_F\ran,\label{a1.2}\ee
where $\lan YW_F\ran = \frac14 tr [ \Gamma (m_1 - i \hat p_1) \bar
\Gamma (m_2 - i\hat p_2 ) \lan W_F\ran_A ]$, and $W_F \equiv \llan
W \rran$, given in (\ref{32}); the spin operator ordering in
(\ref{a1.2}) is not written explicitly. Neglecting spin
dependence, one has a purely scalar function $W_F$, which is
proportional to  a unit $(4\times 4)$ matrix.

Introducing the effective energies $\omega_i = \frac{T}{2s_i}, ~ T
\equiv |x_4 - y_4|$, one can rewrite (\ref{a1.1}) as \be G(x,y) =
\frac{T}{2\pi} \int^\infty_0 \frac{d\omega_1}{\omega_1^{3/2}}
\frac{d\omega_2}{\omega_2^{3/2}} (D^3z^{(1)} D^3z^{(2)}
)_{\vex\vey} e^{-K_1(\omega_1) - K_2(\omega_2)} \llan Y
W_F\rran_{\Delta z_4 },\label{a1.3}\ee and we have taken into
account, that \be \int (Dz^{(1)}_4 Dz_4^{(2)})_{x_4y_4} \lan Y
W_F\ran e^{- \frac 14 \int^{s_1}_0 \left(
\frac{dz^{(1)}_4}{d\tau_1} \right)^2 d\tau_1 - \frac 14
\int^{s_2}_0 \left( \frac{dz^{(2)}_4}{d\tau_2} \right)^2 d\tau_2}=
\frac{\sqrt{\omega_1 \omega_2}}{2\pi T} \llan Y W_F\rran_{\Delta
z_4 }\label{a1.4}\ee

Here $\llan Y W_F\rran_{\Delta z_4 }$ corresponds to the
time-fluctuating Wilson loop average, as in Fig. 1, renormalized
and normalized by the condition \be \llan Y W_F\rran_{\Delta z_4 }
(g=e=0)=1\label{a1.5}\ee
 We omit in what follows the Fock column structure  of the corresponding
 particle  contents in our averaged Wilson loop $\llan Y W_F\rran_{fl}$ with
 the corresponding $Z_i$ factors for each Fock line and concentrate on the
 simplest case of one renormalized closed  $(q_1\bar q_2)$ loop depending on
 $t_E$, as shown in Fig.2. In the neutral case $e_1 =-e_2, \llan Y
 W_F\rran_{\Delta z_4 }$ depends only on coordinate differences $\veta(t_E) = \vez^{(1)}
 (t_E) - \vez^{(2)}(t_E)$, defined at the same moment $t_E$, and one can
 proceed integrating out the c.m. motion. $K_1, K_2$ in (\ref{a1.3}) are
\be K_1 (\omega_1)+ K_2 (\omega_2)  = \left(
\frac{m_1^2+\omega_1^2}{2\omega_1} +
\frac{m_2^2+\omega_2^2}{2\omega_2}\right) T + \int^T_0 d t_E
\left[ \frac{\omega_1}{2} \left( \frac{d\vez^{(1)}}{dt_E}\right)^2
+\frac{\omega_2}{2} \left( \frac{d\vez^{(2)}}{dt_E}\right)^2
\right].\label{a1.5a}\ee

Introducing now the coordinates \be \veta (t_E) = \vez^{(1)} -
\vez^{(2)}, ~~ \verho (t_E) = \frac{\omega_1}{\omega_1+ \omega_2}
\vez^{(1)} (t_E) +\frac{\omega_2}{\omega_1+ \omega_2} \vez^{(2)}
(t_E)\label{a1.6}\ee one can rewrite the last term in
(\ref{a1.5a}) as \be \int^T_0 dt_E \left( \frac{\omega_1+
\omega_2}{2} \left( \frac{ d\verho}{dt_E} \right)^2 + \frac{\tilde
\omega}{2} \left( \frac{ d\veta}{dt_E} \right)^2
\right)\label{a1.7}\ee and the path integral $(D^3z^{(1)}
D^3z^{(2)})_{\vex\vey}$ as

$$(D^3z^{(1)} D^3z^{(2)})_{\vex\vey}= \int \frac{d^3\vep_1}{(2\pi)^3} \int
\frac{d^3\vep_2}{(2\pi)^3} e^{i\vep_1 \left(\sum \Delta \vez^{(1)}
- (\vex-\vey)\right)+ i\vep_2\left (\sum \Delta \vez^{(2)} -
(\vex-\vey)\right)}\times$$\be\times \frac{d^3\Delta
z^{(1)}}{(4\pi \varepsilon_1)^{3/2}}\frac{d^3\Delta z^{(2)}}{(4\pi
\varepsilon_2)^{3/2}}= (D^3 \rho)_{\vex\vey}
(D^3\eta)_{00},\label{a1.8}\ee

Where \be (D^3\verho)_{\vex\vey} = \int\frac{d^3\veP}{(2\pi)^3}
\prod_k e^{i\veP\left(\sum \Delta \verho_k - (\vex-\vey)\right)}
\frac{d^3\Delta \verho_k}{\left(\frac{2\pi \Delta
t_E}{\omega_1+\omega_2}\right)^{3/2}}
 \label{a1.9}\ee

\be (D^3\veta)_{00} = \int\frac{d^3\veq}{(2\pi)^3} \prod_k
e^{i\veq\sum_k \Delta \verho_k } \frac{d^3\Delta
\veta_k}{\left(\frac{2\pi \Delta t_E}{\tilde
\omega}\right)^{3/2}}.
 \label{a1.10}\ee

In absence of external magnetic field, which acts on c.m.
coordinate $\verho$, it is convenient to consider the $\veP=0$
projection of the Green's function
$$\int G(x,y) d^3 (\vex-\vey) = \frac{T}{2\pi} \int^\infty_0
\frac{d\omega_1}{\omega_1^{3/2}}\int^\infty_0
\frac{d\omega_2}{\omega_2^{3/2}}(D^3\veta)_{00} e^{-K(\eta)} \llan
YW_F\rran_{\Delta z_4 }=
$$\be =\frac{T}{2\pi} \int^\infty_0
\frac{d\omega_1}{\omega_1^{3/2}}\int^\infty_0
\frac{d\omega_2}{\omega_2^{3/2}} \lan 0|\lan Y\ran e^{-HT}|0\ran,
\label{a1.11}\ee where

\be  K (\eta)= \left( \frac{m_1^2+\omega_1^2}{2\omega_1} +
\frac{m_2^2+\omega_2^2}{2\omega_2}\right) T + \int^T_0 d t_E
\frac{\tilde \omega}{2} \left(
\frac{d\veta}{dt_E}\right)^2,\label{a1.13}\ee

\be \lan 0|e^{-HT}|0\ran= \sum^\infty_{n=0} |\varphi_n (0) | ^2
e^{- M_n (\omega_1, \omega_2) T}.\label{A1.12}\ee Here
$\left.\varphi_n(0) = \varphi_n (\omega_1, \omega_2,
\veta)\right|_{\veta=0}$, and $M_n (\omega_1, \omega_2)$ is the
eigenvalue of the Hamiltonian \be H\equiv H(\omega_1, \omega_2),
~~ H\varphi_n = M_n (\omega_1, \omega_2)
\varphi_n.\label{a1.13}\ee

Assuming, that $\llan  W_F\rran_{\Delta z_4 }$ can be represented
as \be \llan W_F\rran_{\Delta z_4 } = \exp (- \int \hat V (\eta
,\omega) dt_E),\label{a1.14}\ee the Hamiltonian can be
 written in the
form \be H(\omega_1, \omega_2) = \sum_{i=1,2} \frac{m^2_i +
\omega_i^2}{2\omega_i} + \frac{\vep^2}{2\tilde \omega}+ \hat V
(\eta, \omega_1, \omega_2)\label{a1.15}\ee At this point one can
define the so-called quark decay constants $f_\Gamma^{(n)}$
\cite{30},

\be \int G(x,y) d^3 \vex-\vey) = \sum_n \varepsilon_\Gamma
\bigotimes\varepsilon_\Gamma \frac{\bar M_n (f_\Gamma^{(n)})^2}{2}
e^{-\bar M_n T}\label{A1.16}\ee

where $\varepsilon_\Gamma =1$  for $S$ and $P$ channels, and
$\varepsilon_\Gamma = \varepsilon_\mu^{(k)}$ for $V, A$ channels,
\be \sum_{k=1,2,3} \varepsilon_\mu^{(k)} (q)
\varepsilon_\nu^{(k)} (q) = \delta_{\mu\nu} -\frac{q_\mu
q_\nu}{q^2}\label{a1.17}\ee and hence
 $f_\Gamma^{(n)} $ can be found from (\ref{a1.11}) as

\be (f_\Gamma^{(n)} )^2 e^{-\bar M_n T} = \frac{T}{2\pi}
\frac{2\lan Y\ran}{\bar M_n} \int^\infty_0
\frac{d\omega_1}{\omega_1^{3/2}} \int^\infty_0
\frac{d\omega_2}{\omega_2^{3/2}}\varphi^2_n (0)  e^{-M_n (\omega_1
, \omega_2) T}.\label{a1.18}\ee Here $T$ on both sides is assumed
to tend to $\infty$, and one can  calculate the integral on the
r.h.s. of (\ref{a1.18}) by the stationary point method, $$M_n
(\omega_1, \omega_2) = M_n (\omega_1^{(0)} , \omega_2^{(0)}) +
M_n^{(11)} (\omega_1^{(0)} , \omega_2^{(0)}) \frac{(\omega_1
-\omega_1^{(0)})2}{2}+$$\be+M_n^{(22)} (\omega_1^{(0)} ,
\omega_2^{(0)}) \frac{(\omega_2 -\omega_2^{(0)})2}{2}+M_n^{(12)}
(\omega_1^{(0)} , \omega_2^{(0)}) {(\omega_1
-\omega_1^{(0)})}(\omega_2 -\omega_2^{(0)})\label{a1.19}\ee
 where
  \be  M_n^{(ik)} = \left.\frac{\partial^2 M_n }{\partial \omega_i \partial \omega_k}
\right|_{\omega_i = \omega_i^{(0)},\omega_k =
\omega_k^{(0)}},\label{A1.22}\ee and \be  \left.\frac{\partial M_n
}{\partial \omega_i} \right|_{\omega_i = \omega_i^{(0)}}=0,
i=1,2.\label{A1.23}\ee

Doing the integration in (\ref{a1.18}) with the help of
(\ref{a1.19}) one obtains

$$ (f_\Gamma^{(n)} )^2 = \frac{N_c\lan Y \ran \varphi^2_n (0)}{(\omega_1^{(0)}
  \omega_2^{(0)}) \bar M_n \xi_n}$$
  where $$\xi_n = \sqrt{  \omega_1^{(0)}
\omega_2^{(0)}  \Omega_n},$$ with \be \Omega_n = \frac {\alpha
\beta (\alpha-\beta)^2}{(\alpha -\beta)^2+ \gamma^2} +
\frac{\gamma^2 [(\alpha+ \beta)^2- 2 (\alpha-\beta)^2-\gamma^2]}{4
[(\alpha-\beta)^2 + \gamma^2]}\label{a1.20}\ee where we have
denoted

\be  \alpha = \frac12 M_n^{(11)}, ~~ \beta = \frac12   M_n^{(22)},
~~ \gamma = M_n^{(12)},\label{A1.26}\ee
 and finally \be \bar M_n = M_n (\omega_1^{(0)},
\omega_2^{(0)}), ~~ \bar Y = \frac14 tr_D (\Gamma (m_1- i \hat
p_1) \bar \Gamma(m_2- i\hat p_2)\label{a1.21}\ee
 and $tr_D$ denotes trace over Dirac $4\times 4$ indices.
It is instructive to compare (\ref{a1.20}) with the old result,
obtained in \cite{30}, using approximate path integrals over
$(D\Delta \omega)$, \be (f_\Gamma^{(n)})^2_{\Delta \omega} =
\frac{2N_c \lan Y \ran \varphi_n^2 (0)}{\bar M_n
\omega_1^{(0)}\omega_2^{(0)}}.\label{a1.22}\ee

As one can see, comparing (\ref{a1.20}) and (\ref{a1.22}), in the
first case (the time-fluctuation approach of the present paper)
the factor $\frac{1}{\sqrt{\Omega \omega_1^{(0)}\omega_2^{(0)}}}$
should be equal to 2, for both expressions to coincide. In
practice for the $(q_1\bar q_2)$ state made of zero mass quarks,
$m_1=m_2=0$, and with the total mass made of confining
interaction, see \cite{28} for details, one has \be M_n (\omega_1,
\omega_2) = \sum^2_{i=1} \frac{m_i^2+\omega_i^2}{2\omega_i}+
(2\tilde \omega)^{-1/3} \sigma^{2/3} a_n, a_0 =
2,338\label{a1.23}\ee and for $m_1 =m_2=0$ one obtains \be
(\Omega_0\omega_1^{(0)}\omega_2^{(0)})^{-1/2} =3,\label{a1.24}\ee
while for $m_1=0, m_2\ll \sqrt{\sigma}$, the result is \be
(\Omega_0\omega_1^{(0)}\omega_2^{(0)})^{-1/2} \cong
2.34.\label{a1.25}\ee

This implies, that the quark decay constants $f_\Gamma^{(n)}$
obtained in the new method will be larger by (10-20)\% as compared
with previous calculations in \cite{30}.

\end{document}